\documentclass[10pt,twocolumn,final,journal]{IEEEtran}

\usepackage{bigstrut,multirow,array,psfig,epsfig,subfigure,amssymb,amsmath,color,url,xr}
\usepackage{mathtools}
\usepackage[ruled,linesnumbered]{algorithm2e}
\usepackage{paralist,bbding}
\newcommand{\argmin}{\operatornamewithlimits{argmin}}

\begin{document}
\pagenumbering{arabic}

\title{\emph{RTI Goes Wild}: Radio Tomographic Imaging for\\ Outdoor People Detection and Localization}

\author{Cesare Alippi, \IEEEmembership{Fellow,~IEEE}, Maurizio Bocca, Giacomo Boracchi,\\Neal Patwari\thanks{}, \IEEEmembership{Fellow,~IEEE}\thanks{Cesare Alippi, Maurizio Bocca, Giacomo Boracchi and Manuel Roveri are with the Dipartimento di Elettronica, Informazione e Bioingegneria, Politecnico di Milano, Milan, Italy. E-mail: firstname.lastname@polimi.it. Neal Patwari is with the Department of Electrical and Computer Engineering, University of Utah, and with Xandem Technology, both in Salt Lake City, UT, USA. E-mail: npatwari@ece.utah.edu} and Manuel Roveri}
\maketitle

\begin{abstract}
RF sensor networks are used to localize people indoor without requiring them to wear invasive electronic devices. These wireless mesh networks, formed by low-power radio transceivers, continuously measure the received signal strength (RSS) of the links. Radio Tomographic Imaging (RTI) is a technique that generates 2D images of the change in the electromagnetic field inside the area covered by the radio transceivers to spot the presence and movements of animates (\emph{e.g.}, people, large animals) or large metallic objects (\emph{e.g.}, cars). Here, we present a RTI system for localizing and tracking people outdoors. Differently than in indoor environments where the RSS does not change significantly with time unless people are found in the monitored area, the outdoor RSS signal is time-variant, \emph{e.g.}, due to rainfalls or wind-driven foliage. We present a novel outdoor RTI method that, despite the nonstationary noise introduced in the RSS data by the environment, achieves high localization accuracy and dramatically reduces the energy consumption of the sensing units. Experimental results demonstrate that the system accurately detects and tracks a person in real-time in a large forested area under varying environmental conditions, significantly reducing false positives, localization error and energy consumption compared to state-of-the-art RTI methods.
\end{abstract}


\section{Introduction} \label{sec:introduction}

\PARstart{I}n this paper, we consider the problem of detecting, localizing and tracking people not wearing/ carrying an electronic device, thus not actively participating in the localization effort, in large and heavily obstructed outdoor environments. To this purpose, we use an RF sensor network, \emph{i.e.}, a wireless system composed of low-power, inexpensive commercial of-the-shelf radio transceivers operating in the $2.4$ GHz ISM band \cite{RFSensNet_Proc_IEEE}. These devices form a wireless mesh network by continuously broadcasting packets and measuring the received signal strength (RSS) of the links of the network. Radio tomographic imaging (RTI) \cite{RTI_Wilson_TMC_2010} is applied to process the RSS measurements collected by the RF sensors on multiple frequency channels \cite{ChannelDiv_MASS_2012} and estimate 2D images of the change in the electromagnetic field of the deployment area due to the presence and movements of animates (\emph{e.g.}, people, large animals) or metallic objects (\emph{e.g.}, cars \cite{roadside_surv}). By further processing these images, the targets can be accurately detected and tracked \cite{MTT_Bocca_2013}. Our work introduces an accurate and energy-efficient RTI method that successfully deals with the nonstationary noise introduced in the RSS measurements by environmental factors, such as rainfalls or wind-driven foliage, tipically encountered in real-world outdoor environments.

RF sensor networks represent an appealing technology for detecting and tracking people outdoors: due to the small size and low cost of the RF units, they are less invasive and better conceivable than video camera networks, and considerably less expensive than ultra-wideband (UWB) transceivers. They also work in the dark and through smoke and non-metallic walls. However, an RF sensor network deployed outdoors faces challenging conditions, different from the more static ones typically found in real-world indoor environments \cite{Grandma_2012}. The main challenges encountered in outdoor environments are:

\begin{itemize}
\item The presence of what we refer to as \emph{environmental noise}, \emph{i.e.}, the significant variation in RSS observed when no person is located in the monitored area due to the time-varying multipath introduced by \emph{e.g.} wind-driven foliage, rainfalls or snow. In homes and buildings, on the contrary, the RSS measured in quasi stationary conditions does not vary significantly with time: the spurious variation observed in stationary conditions can be introduced by overlapping Wi-Fi networks increasing the floor noise level of the radio channel \cite{srinivasan2006}.
\item The lack of an initial calibration of the system performed in stationary conditions. Due to the intrinsic nonstationarity of the monitored environment, RTI methods to be applied for outdoor localization can not rely on an initial calibration to estimate the average RSS of the links of the network, which is then used as a reference to calculate the change in the electromagnetic field introduced by the presence of people in the monitored area.
\item The need to address the energy efficiency of the system. The remote locations and harsh weather conditions drive the need for a battery powered capability of the RF sensors, which, in turn, requires the adoption of an adaptive radio duty cycling mechanism to extend the lifetime of the system or, when energy harvesting systems are enforced \cite{Basagni_NRGHarvesting}, reduce their cost and size. 

\end{itemize}

Several application scenarios could take advantage of an RF sensor network for outdoor people detection and localization. For example, in natural parks populated by protected animal species, a system composed of small and covert devices could be used to detect and locate poachers even in areas characterized by a dense canopy, which can not be monitored \emph{e.g.} with air surveillance \cite{drones_poachers}. An outdoor RTI system could be used to monitor long international borders running across heavily forested areas from people trying to \emph{e.g.} smuggle drugs or weapons, to protect large vineyards from vandalism or theft, or to create a virtual fence at a pasture site \cite{Corke_VirtualFence} that would not require the cattle to wear GPS collars. Besides security and perimeter surveillance applications, an RTI system would greatly extend the capabilities of \emph{smart spaces} such as university campuses, malls and hospitals.

In this work, we introduce a novel RTI method for outdoor environments achieving high detection and localization accuracy and improving the overall system's energy efficiency. The main contributions of our work can be summarized as follows:
\begin{itemize}
    \item We characterize the relationship existing in nonstationary outdoor environments between the variation in RSS due to environmental noise and the link's \emph{fade level} \cite{wilson11fade}, \emph{i.e.}, the difference between the measured and theoretical RSS of a link.
    \item We propose a method for selecting those link-frequency channel combinations (which we refer to as \emph{link-channel pairs}) that, even in nonstationary environmental conditions, appear to be the most reliable for detecting the presence of a person on the link line, \emph{i.e.}, the imaginary straight line connecting transmitter and receiver. We verify that this selection method allows enhancing both the detection and localization accuracy while simultaneously reducing the system's energy consumption.
    \item We introduce an adaptive method to recalibrate on-line the reference RSS of the selected link-channel pairs, even in presence of environmental noise and people in the monitored area, and apply background subtraction on the estimated radio tomographic images to further increase the robustness of the system to the time-varying environmental noise.
\end{itemize}

The performance of the outdoor RTI method was evaluated in a set of experiments carried out in a challenging outdoor environment, \emph{i.e}, a $35m \times 60m$ heavily forested area with trees and bushes of various height, shape and size. Tests were performed in different environmental conditions (\emph{e.g.}, with no wind, light breeze, or gusts of wind). False alarm rate, localization accuracy and energy efficiency associated with the novel method are compared to those of the RTI methods originally introduced in \cite{MTT_Bocca_2013,MultiScale_2013}, to date the most accurate RTI methods used in indoor environments, which we suitably adapt to the considered outdoor scenario in order to make a fair comparison. Experimental results demonstrate that the novel outdoor RTI method keeps under control the false positive and negative rates ($0.04$\% and $0$\%, respectively), and reduces both the localization error (from $20$\% to $46$\%) and the energy consumption of the whole system (from $62$\% to $87$\%) compared to state-of-the-art methods.

The paper is organized as follows. In Section \ref{sec:related_literature}, we survey the related literature on people detection and localization in outdoor environments and on state-of-the-art RTI methods. The novel outdoor RTI method is presented in Section \ref{sec:multichannel_RTI}, while the experimental setup is described in Section \ref{sec:setup}. Section \ref{sec:eval_metrics} lists the metrics used to evaluate the performance of the system. The results of the tests are presented in Section \ref{sec:results}. Conclusions are given in Section \ref{sec:conclusion}.

\section{Related Literature} \label{sec:related_literature}

The problem of localizing and tracking people in large outdoor areas has been addressed by several works. The system in \cite{RFIDs_system} was composed of spatially distributed RFID readers measuring the RSS of RFID tags carried by people moving in the monitored area. The position of the targets was estimated by using a RSS-distance model whose parameters were calibrated on-line to make the system more robust to environmental changes. The work in \cite{CamerasRFID_system} presented a heterogeneous system composed of low-quality wireless camera nodes devoted to people localization and RFID readers devoted to identification. Both the systems in \cite{RFIDs_system} and \cite{CamerasRFID_system} assumed that the targets to be located were carrying an RFID tag. Device-free localization and tracking of people has often been carried out by using visual sensor networks \cite{VisSensNet_2009}, \emph{i.e.}, wireless systems composed of a large number of low-power camera nodes. However, cameras produce a large amount of image data for the limited network's resources. In addition, cameras suffer from poor light conditions and occlusions in cluttered outdoor environments (such as \emph{e.g.} forested areas).
For these reasons, other works have proposed using radars to detect personnel in heavily wooded areas. In \cite{Tahmoush_2009,Tahmoush_2013}, micro-Doppler signals, \emph{i.e.}, Doppler scattering returns not due to gross translation of the target but are instead produced by the periodic movements (\emph{e.g.}, of legs or arms) of a walking human target, are exploited for detecting humans and distinguishing them from other animals. However, stand-alone low-frequency radars have a poor angular resolution. To overcome this limitation, the work in \cite{MIMO_radar} proposes a multiple-input multiple-output (MIMO) radar exploiting the angular diversity of spaced antennas to detect the changes in the RF channel due to personnel in the deployment area. In this work, we present an energy-efficient RTI method that enables device-free detection, localization and tracking of people in large and cluttered outdoor areas covered by a small number of low-cost and low-power RF sensors.

The impact of environmental factors on wireless sensor networks deployed outdoors has been previously investigated. The works in \cite{bannister2008wireless,DISI_outdoor_2013} analyze how the density and seasonal variations of vegetation, and daily temperature/humidity fluctuations affect the RSS and the links connectivity. The work in \cite{wind_radio_2006} studies how wind, \emph{i.e.}, wind-driven vegetation movement, affects the propagation of radio signals at different frequencies ($0.9$, $2$, $12$, and $17$ GHz). Experimental data demonstrate that the variation in RSS increases with the wind speed (more consistently at higher frequencies) and that the fade distribution goes from being Rician to being Rayleigh with an increasing wind speed. However, these works did not consider the combined effect of environmental factors and animates on the measured RSS.

Different RTI methods were introduced in \cite{RTI_Wilson_TMC_2010,wilson10see,Kaltiokallio_RTCSA_2011,Zhao_kernel_2013}. In \cite{ChannelDiv_MASS_2012}, frequency diversity was exploited to improve the localization accuracy. The same principle was later applied in \cite{MTT_Bocca_2013,MultiScale_2013}. In all these works, the RTI systems were deployed in stationary indoor environments (\emph{i.e.}, no changes in the environment nor time-varying environmental noise). The system described in \cite{Grandma_2012} was deployed in a real-world domestic environment over a three months period. To recalibrate in an on-line fashion the reference RSS values of the links of the network, the RSS signals were low-pass filtered. By doing this, the system maintained an high localization accuracy ($0.3$ m approximately) despite the frequent changes in the environment. In \cite{Zhao_noise_2011}, the authors carried out tests in a through-wall scenario having some of the RF sensors located in the proximity of trees whose branches were swaying in the wind. They proposed a subspace decomposition method to separate the change in RSS due to human motion from that introduced by environmental noise. However, the amount of variation in the RSS signals due to the environment was estimated during an initial calibration performed with no person in the deployment area. Thus, the method assumed that the spatial and temporal characteristics of the environmental noise were constant. Such an assumption can not be made in a forested outdoor environment, where the environmental noise is highly time-varying.

The RTI methods described above did not consider the issue of energy efficiency: the nodes radio, \emph{i.e.}, the most energy-hungry component, was always on, being the nodes in receive mode at all times, except for when they were transmitting a packet. Energy efficient methods were presented in \cite{kanso09b,Kaltiokallio_RTCSA_2011,Khaledi_SECON_2014}. In \cite{kanso09b}, the authors apply compressed sensing techniques to RTI, reducing the number of links that have to be sampled in order to reconstruct the whole image. In \cite{Kaltiokallio_RTCSA_2011}, an accurate time synchronization protocol is used to enable radio duty cycling. In \cite{Khaledi_SECON_2014}, only the links near the current location of the tracked targets are measured. In our work, we present a method to select those link-channel pairs that are robust to environmental noise (thus, the most informative and reliable for RTI). This approach allows increasing the energy efficiency of the system while simultaneously improving its detection and localization performance.

\section{Outdoor RTI} \label{sec:multichannel_RTI}
This section describes the novel outdoor RTI method. We provide the basics of RTI, focusing on the solutions developed to address the challenges posed by multipath-rich and time-varying outdoor environments. The reader is invited to refer to \cite{RTI_Wilson_TMC_2010,ChannelDiv_MASS_2012,MultiScale_2013,wilson10see,patwari08b,agrawal09} for a detailed description of the principles of RTI. Algorithm \ref{alg:outdoor_RTI_algorithm} details the novel outdoor RTI method.

\begin{algorithm}[t!]
    \SetKwInOut{Input}{input}
    \SetKwInOut{Output}{output}
    
    \Input{$N$ static RF sensors located at (estimated) positions $\{x_n,y_n\}_{n=1,...,N}$\newline Communicating on a set of different frequency channels $\mathcal{C}$\newline Measuring the RSS on the link-channel pairs of the network.}
    \Output{$\hat{\mathcal{P}}(k)$ (estimated positions of the people found in the monitored area at time $k$)}
    
    Calculate the projection matrix $\mathbf{\Pi}$\newline
    \While{(1)}{
        \textbf{At the completion of each full TDMA cycle:}\newline{
            Update the reference RSS $\bar{r}_{l,c}(k)$ of the selected link-channel pairs\newline
            Estimate the change in RSS $y_l$ of the selected link-channel pairs\newline
            Estimate the radio tomographic image $\mathbf{\hat{x}}$ = $\mathbf{\Pi}$$\mathbf{y}$\newline
            Apply background subtraction to $\mathbf{\hat{x}}$ and update the background image\newline
            Apply target detection and tracking method to estimate $\hat{\mathcal{P}}(k)$}
            
        \textbf{Every $\mathbf{\Delta}\mathbf{T_N}$ hours:} (\emph{e.g.}, $\mathbf{\Delta}\mathbf{T_N}$ = 2)\newline{
            Measure the RSS of all link-channel pairs $\{l,c\}$ of the network for $\Delta T_c$ minutes (\emph{e.g.}, $\Delta T_c$ = 5)\newline
            Estimate fade level, $F_{l,c}$, and RSS variance, $\sigma^2_{l,c}$, of the link-channel pairs\newline
            Select the most reliable link-channel pairs to be used for RTI}
    }
\caption{Outdoor RTI method}\label{alg:outdoor_RTI_algorithm}
\end{algorithm}

\subsection{Links Characteristics in Nonstationary Environments} \label{fadeLevel_Noise}

\begin{figure*}[t]
    \begin{center}
        \mbox{
            \subfigure[]{\epsfig{figure=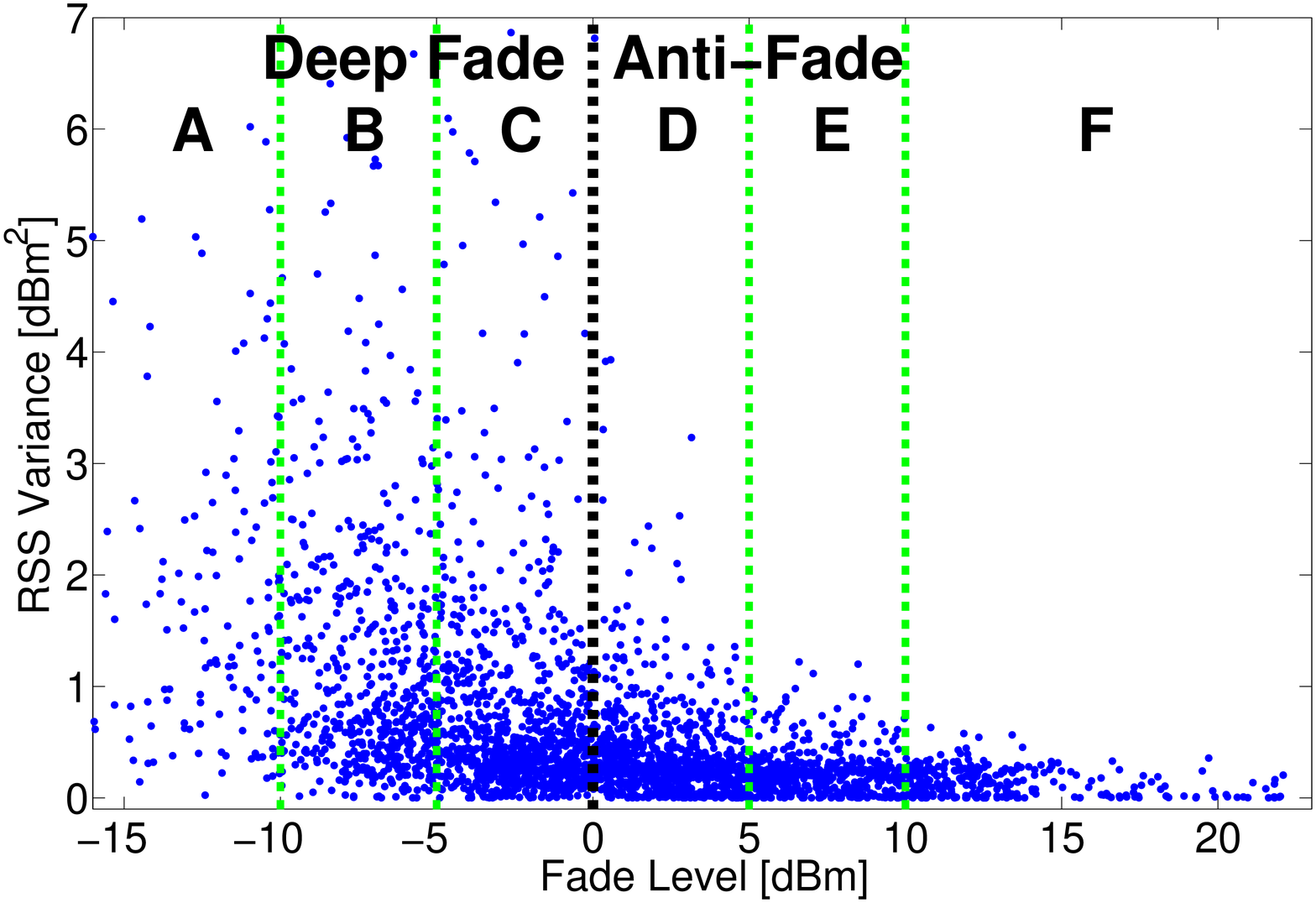,width=\columnwidth}} \quad
            \subfigure[]{\epsfig{figure=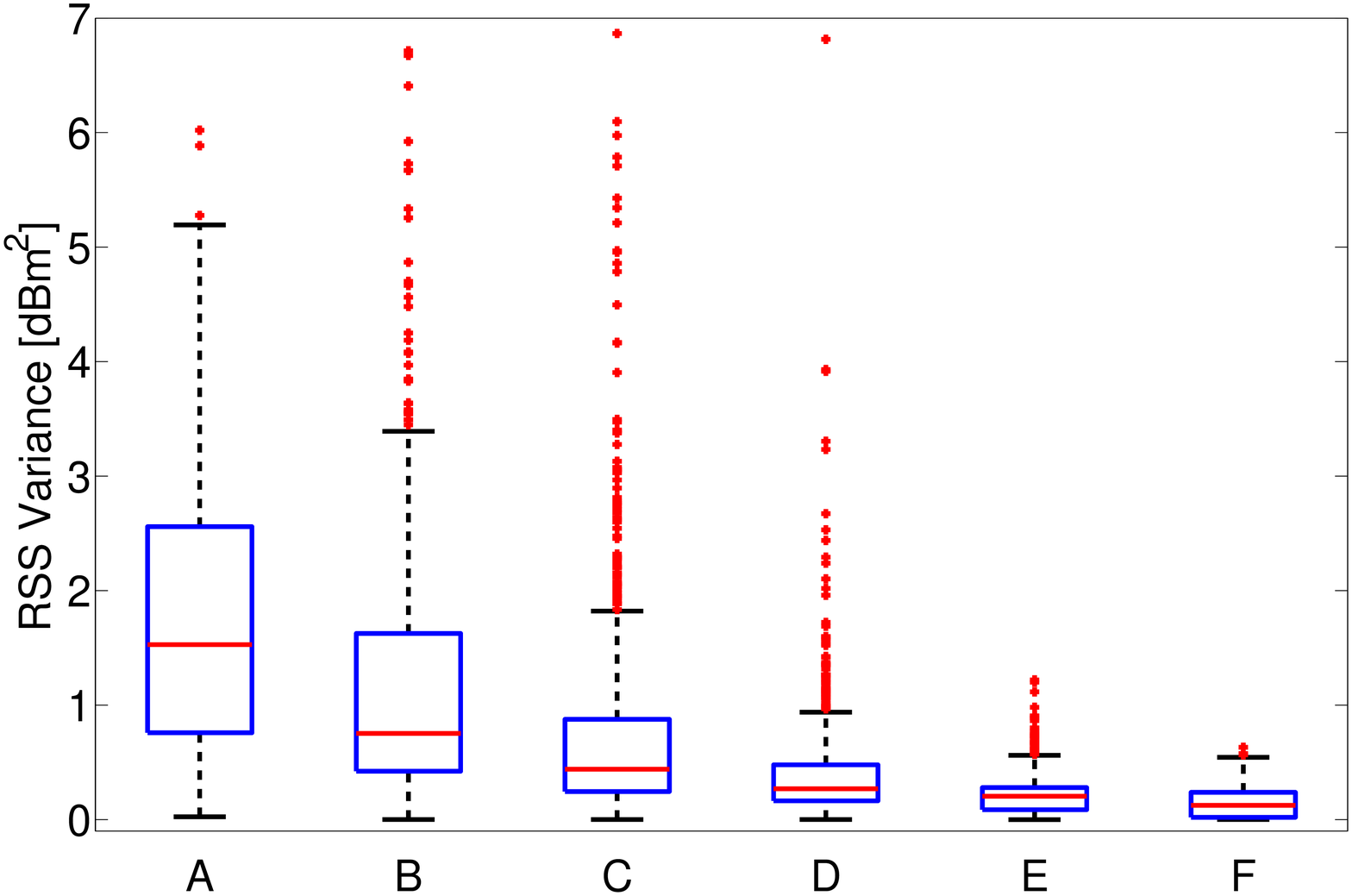,width=\columnwidth}}
        }
        \caption{Fade level \emph{vs.} RSS variance due to environmental noise. The points in Figure (a) are obtained by measuring the RSS of all the link-channel pairs of the network during four intervals of $15$ minutes with wind of varying intensity blowing in the deployment area. Figure (b) shows the boxplots of the distribution of the RSS variance of the link-channel pairs divided in sub-groups (A-F in Figure (a)) based on their fade level. On each box, the central bar represents the median of the distribution, the edges are the $25$th and $75$th percentiles, and the external poinnts are meant to identify outliers, which are plotted individually.}
        \label{fig:FL_var_stationary}
    \end{center}
\end{figure*}

First, we analyze the characteristics of the links in nonstationary outdoor environments. To do this, we use the concept of fade level introduced in \cite{wilson11fade}, which defines the relationship between steady-state, narrow-band fading and the changes in RSS due to a person crossing the link line. The fade level of link $l$ on channel $c$, $F_{l,c}$, can be estimated as:
\begin{equation} \label{eq:fade_level}
    F_{l,c} =  \bar{r}_{l,c} - P(d_l),
\end{equation}
where:
\begin{equation} \label{eq:Pd}
    P(d_l) =  P_0^n - 10 \eta_n \log_{10} \left(\frac{d_l}{d_0^n}\right).
\end{equation}
In (\ref{eq:fade_level}), $\bar{r}_{l,c}$ is the average RSS of link $l$ on channel $c$ measured with no person in the proximity of the link line, and $P(d_l)$ is the theoretical RSS, predicted by using the log-distance path loss model \cite{rappaport96}, for two nodes at distance $d_l$. The equation in (\ref{eq:Pd}) represents a \emph{node-specific} log-distance path loss model whose parameters $\eta_n$, $P_0^n$ and $d_0^n$ (\emph{i.e.}, the path loss exponent, reference path loss, and reference distance, respectively), are derived by fitting the average RSS of those links having node $n$ as transmitter. In other works, \emph{e.g.}, \cite{MultiScale_2013}, a global path loss exponent was estimated by fitting the average RSS of all the links of the network. In this work, by generating for each RF sensor an individual distance-RSS model, hardware variability factors (such as \emph{e.g.} antenna impedance matching or relative antenna orientation between transmitter and receiver \cite{HW_RSS_var_2006}) and local environmental differences (such as \emph{e.g.} the proximity to the node of dense foliage) are taken into consideration \cite{Vallet_model}. In Section \ref{sec:results}, we show that, by deriving an individual model for each node instead of a global one, the system achieves better detection and localization performance.

The fade level can be interpreted as a measure of whether a link-channel pair is experiencing destructive or constructive multipath interference (or not): in the first case, the fade level is negative and the link-channel pair is said to be in \emph{deep fade}; in the latter case, the fade level is positive and the link-channel pair is said to be in \emph{anti-fade}. However, even in multipath-rich indoor environments, the measured RSS does not vary significantly unless, \emph{e.g.}, a human body affects the propagation of one (or more) of the multipath components. On the contrary, outdoor environmental factors, such as wind, rainfall or snow, introduce a significant variation in RSS even when no person is found in the deployment area. We measure the effect of the environmental noise on a link-channel pair as the variance of the RSS measurements collected when no person is in the proximity of the link line. The relationship between fade level and environmental noise is depicted in Figure \ref{fig:FL_var_stationary}. The points in Figure \ref{fig:FL_var_stationary}(a) are obtained by measuring the RSS of all the link-channel pairs of the system described in Section \ref{sec:setup} during four different intervals of $15$ minutes with wind of varying intensity blowing in the deployment area. Figure \ref{fig:FL_var_stationary}(b) shows the boxplot of the RSS variance of all the link-channel pairs, divided in six sub-groups (A-F) based on their fade level. The data show that link-channel pairs in deep fade, \emph{i.e.}, those having negative fade level, have a higher RSS variance when the wind becomes stronger. Thus, those links hardly contribute to the detection and localization problems.

\begin{figure}[t]
    \begin{center}
        \epsfig{figure=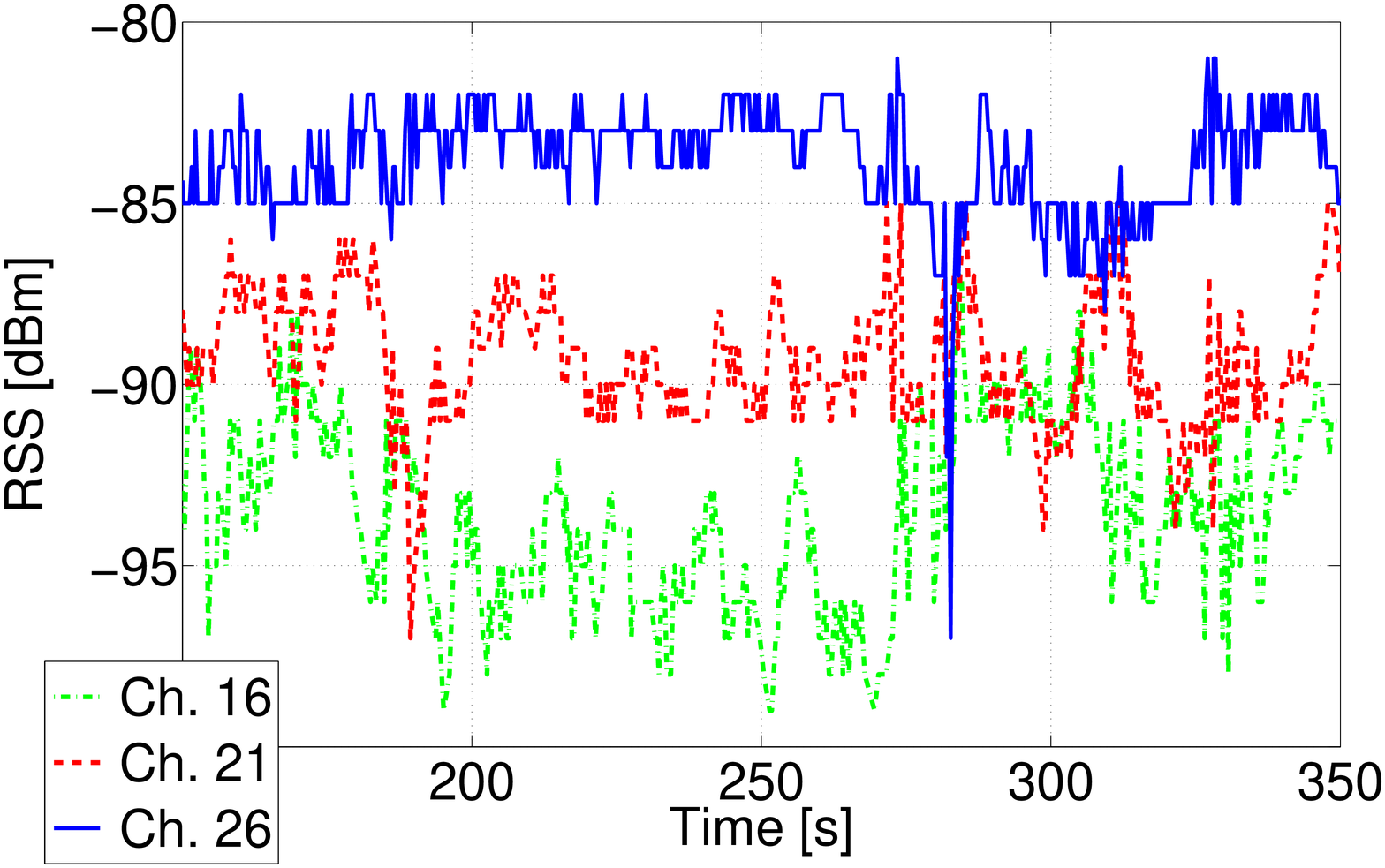,width=\columnwidth}
        \caption{The RSS, measured on different frequency channels, of a $25$ m link cutting through several branches of trees found in the area of the test. During the test, carried out in a day with wind of moderate intensity, a person crossed the link line at $t = 282$ s.}
        \label{fig:link_example}
    \end{center}
\end{figure}

Data in Figure \ref{fig:FL_var_stationary} show that link-channel pairs with a positive fade level (\emph{i.e.}, in anti-fade) are more robust to environmental noise. Moreover, link-channel pairs in anti-fade are characterized by a smaller sensitivity area \cite{wilson11fade,MultiScale_2013}, \emph{i.e.}, the area where a person affects the RSS. This area is modeled as an ellipse having transmitter and receiver at the foci \cite{patwari08b}: for link-channel pairs in anti-fade, it can be modeled as a \emph{narrow} ellipse around the link line. Thus, link-channel pairs with a positive fade level are not only more robust to environmental noise but can also be considered more reliable indicators of the presence of a person in the proximity of the link line. We provide an example of this in Figure \ref{fig:link_example}, which shows the RSS, measured on different frequency channels, of a $25$ m link cutting through several branches of the trees found in the area. During the test, carried out in a day with wind of moderate intensity, a person crossed the link line at $t = 282$ s. The only channel in anti-fade, \emph{i.e.}, channel $16$ having $F_{l,c} = 2.1$ dBm, measured a consistent attenuation ($12$ dBm) in RSS at the crossing event, while exhibiting small variation in RSS due to the movements of the wind-driven foliage. On the contrary, the other two channels in deep fade show a continuous, significant variation in RSS due to the action of the wind, and this does not allow to unequivocally detect the crossing event.

\subsection{Link-Channel Pairs Selection} \label{links_sel_weighting}

We now present a method to select a subset of link-channel pairs to be used for RTI. The reason for doing this is twofold: on one hand, we want to select link-channel pairs that are more robust to environmental noise and, at the same time, more informative about the position of the targets in order to enhance the detection and localization performance of the system. On the other, we want to increase the energy efficiency of the system, since the radio of the RF sensors can be turned off during the TDMA slots originally allocated to the discarded link-channel pairs.

Define $\mathcal{L}_p$ to be the set of link-channel pairs with positive fade level
\begin{equation} \label{eq:anti-fade_set}
    \mathcal{L}_p =  \left\{ \left( l,c \right) : F_{l,c} > 0 \right\},
\end{equation}
where $l$ is the link and $c$ the frequency channel. 

Due to the low nodes density typical of outdoor deployments and the presence of several obstructions, some link-channel pairs may measure a RSS close to the sensitivity threshold of the radio modules ($-97$ dBm at typical ambient temperatures for the radios described in Section \ref{hardware_protocol}). Links with an average RSS close to the sensitivity threshold belong to a \emph{grey region} \cite{RSSI_underappreciated} in which their connectivity performance becomes highly unpredictable and their RSS measurements include noise and interference due to very weak signals. Moreover, the connectivity of these links is highly affected by changes in the environment \cite{Ceriotti_jungle}. Thus, we discard all link-channel pairs with an average RSS lower than a pre-defined threshold $\Upsilon_r$ (we set $\Upsilon_r = -90$ dBm). 

Define $\mathcal{L}_r$ to be the set of link-channel pairs with average RSS higher than $\Upsilon_r$
\begin{equation} \label{eq:high_average_set}
    \mathcal{L}_r =  \left\{ \left( l,c \right) : \bar{r}_{l,c} > \Upsilon_r \right\}.
\end{equation}
Now define $\mathcal{L}_s = (\mathcal{L}_p \cap \mathcal{L}_r)$ to be the set of link-channel pairs to be considered. Both the fade level and the average RSS of the link-channel pairs are estimated when no person is found in the deployment area.

For each link-channel pair, we calculate a weight $\rho_{l,c}$ as follows
\begin{equation}\label{eq:weight_link_channel}
    \rho_{l,c} = \begin{cases}
        F_{l,c} / \sigma^2_{l,c} & \text{if } \left( l,c \right) \in \mathcal{L}_s\\
        0 & \text{otherwise}
    \end{cases},
\end{equation}
where $\sigma^2_{l,c}$ is the RSS variance measured when no person is found in the deployment area. Consequently, link-channel pairs having low RSS variance are assigned a higher weight, since the variations in RSS measured on such links have a higher probability of being human-induced.

Finally, define the set $\mathcal{L}$ as
\begin{equation} \label{eq:final_set}
    \mathcal{L} =  \left\{ \left( l,c \right) : \left( l,c \right) \in \mathcal{L}_s \land c = \max_{j \in \mathcal{C}} \rho_{l,j} \right\},
\end{equation}
\emph{i.e.}, as the set in which, for each link in $\mathcal{L}_s$, only the frequency channel in $\mathcal{C}$ (where $\mathcal{C}$ represents the set of measured frequency channels) characterized by the largest weight is included.

\begin{figure}[t]
    \begin{center}
        \epsfig{figure=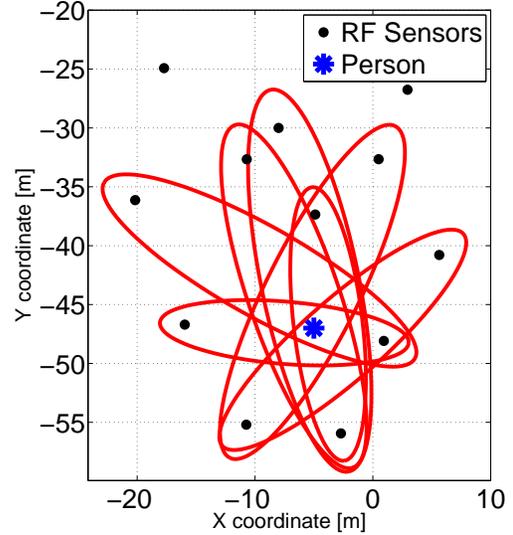,width=0.8\columnwidth}
        \caption{At time $k$, the FIFO buffers of the selected link-channel pairs whose sensitivity areas (red ellipses) include a person's estimated position are not updated. The FIFO buffers of the other selected link-channel pairs are instead updated. For each link-channel pair in $\mathcal{L}$, the reference RSS is calculated as the average of the RSS measurements contained in its FIFO buffer.}
        \label{fig:sens_area}
    \end{center}
\end{figure}

\subsection{RSS Change Estimation} \label{online_RSS_change}

The change in RSS at time $k$ for each link-channel pair in $\mathcal{L}$ is estimated as:
\begin{equation} \label{eq:RSS_change}
    y_{l,c} =  | r_{l,c}(k) - \bar{r}_{l,c}(k) |,
\end{equation}
where $r_{l,c}(k)$ and $\bar{r}_{l,c}(k)$ are the \emph{measured} and \emph{reference} RSS of the link-channel pair $(l,c)$, both considered at time $k$, respectively. In most of the previous works (\emph{e.g.}, \cite{RTI_Wilson_TMC_2010,MTT_Bocca_2013,MultiScale_2013}), the reference RSS was calculated as the average RSS measured during an initial calibration of the system carried out in absence of people in the monitored area. This approach showed to be effective in time invariant indoor environments. In \cite{Grandma_2012}, the reference RSS was estimated in real-time by low-pass filtering the measured RSS. This approach was effective in a dynamic domestic environment characterized by frequent, bursty events modifying the attenuation field of the deployment area. However, this approach was not effective when the person to be localized was not moving for an extended period of time (\emph{e.g.}, sleeping): in that case, the reference RSS quickly converged to the measured RSS, bringing the corresponding change in RSS to zero despite the presence of the person in the deployment area. Due to the presence of a time-varying environmental noise affecting the RSS measurements, an outdoor RTI system must continuously update the reference RSS of the selected link-channel pairs, even when people are not moving inside the deployment area.

To estimate the reference RSS in real-time without losing people who stop moving, we use the following procedure. We first create, for each selected link-channel pair, a FIFO buffer of $N_w = \lfloor {T_w}/{T_s} \rfloor$ elements. $T_w$ is the length of the considered time window (\emph{e.g.}, $T_w = 5$ s) and $T_s$ is the sampling interval of the RTI system, \emph{i.e.}, the interval of time required by the system to complete all the TDMA rounds of communication on the frequency channels in $\mathcal{C}$ (see Section \ref{hardware_protocol}). Define $\mathcal{\hat{P}}(k)$ to be the set of estimated positions at time $k$ of the people located in the deployment area ($\mathcal{\hat{P}}(k) = \emptyset$ if no person is in the deployment area). Since the change in RSS of a link is assumed to be a spatial integral of the attenuation field of the monitored area, only the attenuation in the elliptical sensitivity area of the link will affect its RSS. Thus, at time $k$, only the FIFO buffers of link-channel pairs whose sensitivity areas do not include one of the positions in $\mathcal{\hat{P}}(k)$ are updated. For each link-channel pair in $\mathcal{L}$, the reference RSS $\bar{r}_{l,c}(k)$ is calculated as the average of the RSS measurements contained in its FIFO buffer. In this way, the system does not require an initial calibration to calculate the reference RSS. Moreover, the reference RSS does not converge to the measured RSS even when the person stops moving for an extended period of time.


\subsection{Radio Tomographic Image Estimation} \label{image_estimation}

When we consider all the $N(N-1)$ links of the network, the change in the electromagnetic field of the deployment area $\mathbf{x}$ is estimated as:
\begin{equation} \label{eq:matrix_RTI}
    \mathbf{y} = \mathbf{W}\mathbf{x} + \mathbf{n},
\end{equation}
where $\mathbf{y}$ and $\mathbf{n}$ are vector of size $N(N-1)$ of the RSS change and noise of the links of the network, respectively. For the links belonging to $\mathcal{L}$, the RSS change is calculated as in (\ref{eq:RSS_change}). For all the others, the RSS change is zero. Vector $\mathbf{x}$ represents the change in the \emph{discretized} electromagnetic field of the monitored area, \emph{i.e.}, the intensity of each pixel of the radio tomographic image to be estimated. Thus, $\mathbf{x}$ is a vector of size $P$, where $P$ is the total number of pixels of the radio tomographic image. Element $w_{l,q}$ of the weight matrix $\mathbf{W}$ indicates how pixel $q$ affects the change in RSS of link $l$. Since the sensitivity area of a link is modeled as an ellipse \cite{patwari08b}, $w_{l,q}$ is computed as:
\begin{equation} \label{eq:weight model}
    w_{l,q} = \begin{cases}
        \frac{1}{A_l} & \text{if } d_{l,q}^{TX}+d_{l,q}^{RX}<d_{l}+\lambda\\
        0 & \text{otherwise}
    \end{cases},
\end{equation}
where $A_l$ is the area of the ellipse, \emph{i.e.}, the sensitivity area, of link $l$, $d_{l,q}^{TX}$ and $d_{l,q}^{RX}$ are the distances of pixel $q$ from the transmitter and receiver, respectively, and $\lambda$ is the parameter defining the width of the sensitivity area. The linear model in (\ref{eq:matrix_RTI}) is based on the correlated shadowing models in \cite{RTI_Wilson_TMC_2010,patwari08b,agrawal09}.

Regularization \cite{Vogel_book} is required to solve the ill-posed problem of estimating the intensity of the \emph{many} pixels in $\mathbf{x}$ from the \emph{few} links' measurements in $\mathbf{y}$. We apply the regularized least square approach used also in \cite{MTT_Bocca_2013,MultiScale_2013}:
\begin{equation} \label{eq:linear_transformation}
    \hat{\mathbf{x}} =  \mathbf{\Pi}\mathbf{y}.
\end{equation}
where:
\begin{equation} \label{eq:tikhonov}
    {\mathbf{\Pi}} = {(\mathbf{W}^T\mathbf{W}+\mathbf{C}_{x}^{-1}\alpha_r)}^{-1}\mathbf{W}^T.
\end{equation}
$\alpha_r$ is the regularization parameter (\emph{e.g.}, $\alpha_r = 0.1$). The a priori covariance matrix $\mathbf{C}_{x}$ is calculated using an exponential spatial decay model \cite{patwari08b}:
\begin{equation} \label{eq:cov_matrix}
    \mathbf{C}_{x}[i,j]=\sigma_{x}^{2}e^{-d_{i,j} /\delta_{c}},
\end{equation}
where $\sigma_{x}^{2}$ is the variance at each pixel, $d_{i,j}$ is the Euclidean distance between pixels $i$ and $j$, and $\delta_{c}$ is a pixels' correlation distance parameter. The inversion matrix $\mathbf{\Pi}$, of size $P \times N(N-1)$, has to be calculated only once after the deployment of the RF sensors, so that $\hat{\mathbf{x}}$ in (\ref{eq:linear_transformation}) can be estimated in real-time.

\subsection{Background Subtraction} \label{sec:background_sub}

Background subtraction is often used in machine vision in order to detect moving objects in videos recorded by static cameras. By iterating the process described in Section \ref{image_estimation} and exploiting the temporal continuity of the estimated images, these can be considered as the \emph{frames} of a video in which the targets to be detected and tracked are the blobs corresponding to people entering and moving in the monitored area. Over the years, several background subtraction methods have been proposed \cite{Piccardi_back_sub,Benezeth_back_sub}. Motion detection is performed by calculating the pixel-wise difference between the current frame and a reference (or background) image, which in our case is the estimated electromagnetic field of the monitored area when no person is found in it. Only the pixels whose intensity is significantly different from that of the background image are labeled as \emph{foreground} pixels, thus considered as potential moving objects, \emph{i.e.}, people. In a RTI system deployed outdoor, the background image can change, gradually or suddenly, due to the environmental noise in the RSS signals. Thus, it must be continuously updated to increase the robustness of the motion detection process.

We model each pixel of the background image as a realization of a Gaussian distribution $\mathcal{N}(\mu_p(k),\sigma_p^2(k))$, where $\mu_p(k)$ and $\sigma_p(k)$ are the expectation and standard deviation of the intensity of pixel $p$ at time $k$. These parameters are computed using the corresponding FIFO buffers containing $N_b = \lfloor {T_b}/{T_s} \rfloor$ elements for each pixel. The set of background pixels $\mathcal{B}(k)$ is defined as:
\begin{equation} \label{eq:back_foreground}
    \mathcal{B}(k) = \left\{q : \frac{|\hat{\mathbf{x}}_q(k) - \mu_q(k-1)|}{\sigma_q(k-1)} \le K_b\right\},
\end{equation}
where $K_b$ is a threshold defining the confidence interval of the intensity of the pixels (we set $K_b = 1$) and $\mu_q(k-1)$ and $\sigma_q(k-1)$ are the mean and standard deviation of the Gaussian distribution of pixel $q$ estimated at time $k-1$, respectively. At time $k$, only the FIFO buffers of the pixels in the background are updated with the current intensity value.

At time $k$, the background image $\mathbf{M}(k)$ is calculated as a $P \times 1$ vector whose elements are the mean of the pixels' FIFO buffers:
\begin{equation} \label{eq:background_image}
    \mathbf{M}(k) = \left[ \mu_1(k),...,\mu_P(k) \right].
\end{equation}
Finally, the background subtracted radio tomographic image, $\hat{\mathbf{x}}_b$, is calculated as:
\begin{equation} \label{eq:back_sub_RTI}
    \hat{\mathbf{x}}_b(k) = \hat{\mathbf{x}}(k) - \mathbf{M}(k).
\end{equation}
The background subtraction allows increasing the difference in intensity between the pixels of the blobs corresponding to real people and those of the blobs introduced by the environmental noise, making the blobs tracking more accurate. 


\subsection{People Detection and Tracking} \label{MTT_subsection}

To detect and track the blobs corresponding to real people, we apply on $\hat{\mathbf{x}}_b$ the method in \cite{MTT_Bocca_2013}, providing high accuracy ($0.5$ m tracking error approximately) and real-time performance with multiple people even when these have intersecting trajectories. We refer the reader to \cite{MTT_Bocca_2013} for a detailed description of the applied multiple target tracking method. By applying the same detection and tracking methods, we can fairly evaluate the performance of the novel outdoor RTI method presented in this work and compare it to previous works. The parameters of the method presented in Section \ref{sec:multichannel_RTI} are summarized in table \ref{t:table_parameters}.

\begin{table}[t!]
    \caption{RTI parameters (default values)} 
        \centering
        \footnotesize
        \begin{tabular}{c c c} 
        \hline\hline\          
        Parameter & Value & Description \\
        \hline  
        $\Upsilon_r$     & $-90$     & Link connectivity threshold [dBm]\\
        $T_w$            & $5$       & Link RSS FIFO buffer length [s]\\
        $\lambda$        & $2$       & Link ellipse (sensitivity area) width\\
        $T_b$            & $5$       & Background image training period [s]\\
        $K_b$            & $1$       & Background/Foreground pixel threshold\\
        $\alpha_r$       & $0.1$     & Regularization parameter\\
        $\sigma_{x}^{2}$ & $0.001$   & Pixels' intensity variance\\
        $\delta_{c}$     & $1$       & Pixels' correlation distance\\
        $p$              & $0.65$    & Pixel width [m]\\      
        \hline 
        \end{tabular}
        \label{t:table_parameters}
\end{table}

\section{Experimental Setup} \label{sec:setup}

\subsection{Hardware and Communication Protocol} \label{hardware_protocol}

In our experiments, we use TI CC2531 nodes \cite{TI_CC2531}, equipped with a SWRU120b antenna \cite{antenna_nodes}. The CC2531 has a nominal maximum transmit power of $4.5$ dBm, and can transmit on one of $16$ frequency channels, which are $5$ MHz apart, in the $2.4$ GHz ISM band. Other frequency bands, \emph{e.g.}, $900$ MHz and $433$ MHz, could be similarly used for outdoor RTI. An analysis of the performance of different frequency bands for RF-based people localization in outdoor environments is outside the scope of this paper and is left for future research. The nodes used in the experiments drain approximately $35$ mA when the radio is on, and $20$ $\mu$A when off.

The RF sensors run a multi-channel TDMA communication protocol, described in detail in \cite{RTI_book}, in which each node has a unique slot number, and transmits only during its slot. Differently than in typical indoor environments where the mesh network formed by the RF sensors is fully connected, in outdoor deployments some nodes may not receive packets transmitted by other nodes, and the connectivity of links (particularly those in the grey region \cite{RSSI_underappreciated}) may vary significantly over time. The nodes use the information included in the received packets, \emph{i.e.}, ID of the transmitting node and total number of nodes in the network, to synchronize their TX/RX schedules and synchronously switch on the next frequency channel in $\mathcal{C}$. This mechanism does not require a command from a central coordinating unit, making the system robust to lossy links and nodes' failure. The transmitted packets contain also the RSS of the most recent packets received from the other nodes. In our experimental setup, a sink node listens to all the packets transmitted by the nodes. The sink node is connected to a laptop where the RSS measurements are stored for post-processing.


\subsection{Experiments} \label{experiments}

\begin{figure}[t]
    \begin{center}
        \epsfig{figure=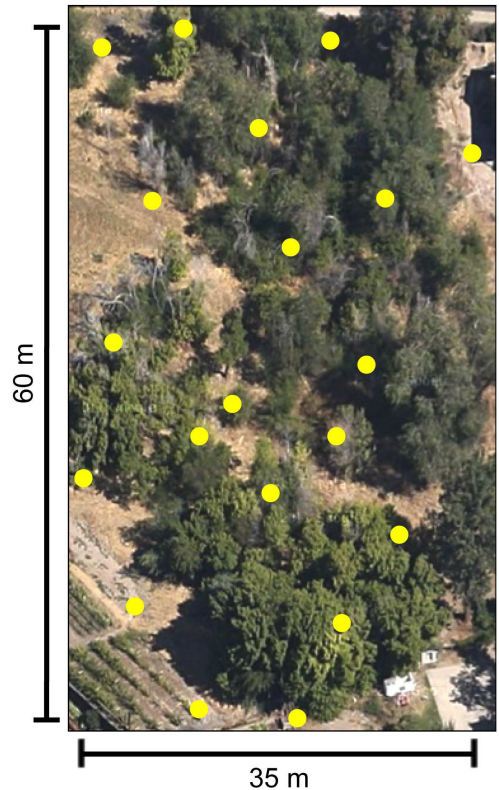,width=0.8\columnwidth}
        \caption{Aerial view of the forested area where the experiments were carried out. The yellow circles represent the $20$ RF sensors composing the RTI system, in the positions estimated by using the procedure described in Section \ref{coord_system}.}
        \label{fig:forest_deployment}
    \end{center}
\end{figure}

We perform experiments in a $35m \times 60m$ heavily forested area, shown in Figure \ref{fig:forest_deployment}. The $20$ battery-powered RF sensors composing the system are attached using tape to the trunks of the trees found in the area, at different heights from the ground. The nodes communicate on four frequency channels, specifically $\mathcal{C} = \{11,16,21,26\}$. The center frequency of a 802.15.4 channel is calculated as $f_c = 2400 + 5(c-10)$ MHz, where $c$ is the channel number. Each selected link-channel pair is measured every $T_s = 0.34$ s (\emph{i.e.}, an approximate sampling frequency of $3$ Hz). To estimate the position of the nodes, we measure at least two length-angle tuples from each sensor. Similarly, we estimate the position of other reference points throughout the deployment area. These points are used to control the movements of the person to be localized and estimate his true position during the tests.

The system run for several hours, during which we collected RSS measurements with no people in the monitored area and carried out several tests in different environmental conditions (\emph{i.e.}, absence of wind, presence of a light breeze, with moderately strong gusts of wind). In each test, a person walks in the deployment area carrying an audio recorder. The person moves along straight lines connecting the reference points and the RF sensors. Each time the person passes nearby a reference point or a sensor, he speaks into the audio recorder its ID number. In addition, the person stands without moving for some time (\emph{e.g.}, 20 s) at various spots inside the deployment area. By interpolating the time stamps in the audio recording, we are able to always derive the true position of the person.

In section \ref{sec:results}, we present the results of tests in which only one person at a time is located in the monitored area. However, the outdoor RTI method presented in this work can be applied to localize and track multiple people.

\subsection{RF Sensors' Position Estimation} \label{coord_system}

A RTI system is composed of $N$ static RF sensors located at estimated positions $\{\hat{p}^n_x,\hat{p}^n_y\}_{n=1,...,N}$. In indoor environments, the availability of a blueprint and structural reference points enables a more precise estimation of the sensors' position. In outdoor environments, the lack of a blueprint and other reference points, the uneven terrain, and the eventual presence of trees and bushes affecting the intra-node distance measuring process increase the error in the sensors' position estimates. In a forested environment, a GPS could be used to locate the RF sensors. However, trees' canopy, trunks and dense foliage near the receiving antenna can interfere with the reception of the signals broadcasted by the satellites, causing huge positional errors, or totally block these signals, making the positioning impossible \cite{GPS_forest,GPS_forest2}.

We now present a method to estimate the position of the RF sensors in a forested environment in which GPS measurements are not available. The $N$ RF sensors are deployed at different heights from the ground, creating a mesh of link lines that conforms to the terrain undulation and capable of intersecting a human body (contrary to what happens in indoor deployments, in this case the sensors are not on the same 2D plane). We choose one sensor to be the reference one, having coordinates $(0,0)$ in the 2D space. We proceed by manually measuring the \emph{length} and the \emph{angle} (relative to the North geomagnetic pole) of a limited number of links (measuring all the links would be incredibly time consuming), making sure that the measured links connect all the deployed sensors.

The initial position estimates are determined by considering only the measurements of links which sequentially connect all the deployed sensors. Starting from the initial position estimates so derived, we iteratively find the maximum likelihood estimate (MLE) of the position of each sensor by taking into consideration also the other measured links. The final position estimates of the sensors are calculated as:
\begin{equation}
    \argmin_{\{\hat{p}^n_x,\hat{p}^n_y\}} \left( \frac{\sum_{l \in \mathcal{T}} \left( d_l - \hat{d}_l \right)^2}{\sigma^2_d} + \frac{\sum_{l \in \mathcal{T}} \left( \alpha_l - \hat{\alpha}_l \right)^2}{\sigma^2_{\alpha}} \right),
\end{equation}
where $\mathcal{T}$ is the set of links manually measured, $d_l$ and $\alpha_l$ are the measured length and angle of the links in $\mathcal{T}$, respectively, and $\hat{d}_l$ and $\hat{\alpha}_l$ are the estimated length and angle of the links in $\mathcal{T}$, respectively, which change at each iteration depending on the estimated positions of the RF sensors. $\sigma^2_d$ and $\sigma^2_{\alpha}$ are the variance of the length and angle measurement process, respectively (we set $\sigma^2_d = 0.5$ m and $\sigma^2_{\alpha} = 5^{\circ}$ after collecting multiple consecutive measurements for the same links). Other methods can be considered to provide an estimate of the spatial coordinates of the sensing units, \emph{e.g.}, see \cite{Alippi_book} for a review.



\section{Evaluation Metrics} \label{sec:eval_metrics}

To evaluate the performance of the outdoor RTI method and of the previous methods here discussed, we use three figures of merit:

\begin{inparadesc}
    \item[Energy efficiency coefficient:]measured as the ratio of link-channel pairs that are not selected to be used for RTI, \emph{i.e.}, the ratio of TDMA slots the nodes remain in energy-saving mode, with the radio off:
    \begin{equation} \label{eq:nrg_eff}
        \vartheta_e = 1- \frac{\left|\mathcal{L}\right|}{N(N-1)\left|\mathcal{C}\right|}.
    \end{equation}

    \item[False alarm rate:]the system triggers a false alarm whenever it wrongly detects at least a person in the monitored area. We measure the false alarm rate as the percentage of radio tomographic images where the real and estimated number of people in the monitored area differ.
    
    \item[Localization accuracy:] measured as the root mean squared error (RMSE) of the position estimates, $\hat{z}(k)$, provided by the RTI system:
    \begin{equation} \label{eq:RMSE}
        \bar{e} = \left( \frac{1}{K} \sum_{k=1}^K{\left( \hat{z}(k)-z(k) \right)}^2 \right)^{1/2},
    \end{equation}
    where $K$ is the total number of position estimates provided by the system during the deployment and $z(k)$ the true position of the person at time $k$. We calculate the RMSE both when the person is moving, $\bar{e}_m$, or standing, $\bar{e}_s$.
\end{inparadesc}

\section{Experimental Results} \label{sec:results}

\begin{center}
\begin{table*}[t!]
    \caption{Summary of the performance of outdoor RTI methods} 
        \centering
        \footnotesize
        \begin{tabular}{c | c c | c c | c c c c}
        \hline\hline 
        &  \multicolumn{2}{c|}{PATH LOSS EXPONENT} & \multicolumn{2}{c|}{LINK-CHANNEL SELECTION} & \multicolumn{4}{c}{PERFORMANCE}\\
        \hline
        METHOD & Node-specific & Global & Heaviest channel & Weighted average & $\vartheta_e$ & False alarm rate [\%] & $\bar{e}_m$ [m] & $\bar{e}_s$ [m]\\
        \hline
        \emph{$OUT^{+}$} & \checkmark &            & \checkmark &            & 87.2 & 0.04 & 3.8 & 3.2\\
        \hline 
        \emph{$OUT^{w}$} & \checkmark &            &            & \checkmark   & 67.2 & 0.07 & 3.7 & 3.2\\
        \hline 
        \emph{$COM^{+}$} &            & \checkmark & \checkmark &            & 88.3 & 0.06 & 5.0 & 4.2\\
        \hline 
        \emph{$COM^{w}$} &            & \checkmark &            & \checkmark   & 68.7 & 0.07 & 4.9 & 4.1\\
        \hline 
        \end{tabular}
        \label{t:RTI_methods_table}
\end{table*}
\end{center}

In this section, we first present the experimental results of the outdoor RTI method. We also consider modified versions of the original algorithm, and demonstrate that the method we propose outperforms the modified versions in terms of false alarm rate, energy efficiency and localization accuracy. Then, we evaluate previous indoor RTI methods, which we suitably adapt to the considered outdoor scenario in order to make a fair comparison. Finally, we analyze the sensitivity of the outdoor RTI method to the parameters in Table \ref{t:table_parameters}.

\subsection{Outdoor RTI Method Performance} \label{new_methods_results}

We evaluate the performance of the outdoor RTI method by post-processing all the RSS measurements collected during the deployment of the system. The system operated for approximately $8$ hours in a day with wind of varying intensity. During this time, the deployment area remained empty approximately $90$\% of times, while in the remaining $10$\% a person was present. The set $\mathcal{L}$ of the link-channel pairs selected and used for RTI is updated every two hours. Thus, $\mathcal{L}$ is updated a total of four times during the deployment. Each time, the system measures the RSS of all the link-channel pairs of the network for $5$ minutes. At the end of this interval, the link-channel pairs of the network are weighted and the most reliable selected.

First, we evaluate the performance of the novel outdoor RTI method described in Section \ref{sec:multichannel_RTI}. We name the method \emph{$OUT^{+}$}. On average, only $12.8$\% of the link-channel pairs are selected, \emph{i.e.}, $\vartheta_e = 87.2$\%. When the deployment area is empty, the RTI system triggers two false alarms, detecting the presence of a person in the monitored area for $12$ s, \emph{i.e.}, a $0.04$\% false alarm rate. Instead, when a person is located in the monitored area area, the RMSE is $\bar{e}_m = 3.8$ m when the person is moving and $\bar{e}_s = 3.2$ m when the person does not move.

\subsection{Modified Outdoor RTI Methods} \label{sec:OUT_modVersions}

We now evaluate the performance of modified versions of the original outdoor RTI method \emph{$OUT^{+}$}. First, we extend the set of selected link-channel pairs $\mathcal{L}$ by considering for each link $l \in \mathcal{L}_s$ the RSS measurements collected on all the selected frequency channels (instead of considering only the frequency channel having the highest weight $\rho_{l,c}$). In this version of the outdoor RTI method (which we name \emph{$OUT^{w}$}), the change in RSS for link $l$ at time $k$ is calculated as:
\begin{equation} \label{eq:weighted_RSSchange}
    y_l(k) = \frac{\sum_{c \in \mathcal{L}_s} \rho_{l,c} | r_{l,c}(k) - \bar{r}_{l,c}(k) |}{\sum_{c \in \mathcal{L}_s} \rho_{l,c}},
\end{equation}
\emph{i.e.}, as the weighted average of the change in RSS measured on the selected frequency channels. This method that merges the RSS data collected for the same link on different frequency channels was originally introduced in \cite{MTT_Bocca_2013}. With \emph{$OUT^{w}$}, the system selects on average $32.8$\% of the link-channel pairs, \emph{i.e.}, $\vartheta_e = 67.2$\%, and has a $0.07$\% false alarm rate (two triggered false alarms, person detected in the deployment area for $18$ s). The RMSEs are $\bar{e}_m = 3.7$ m and $\bar{e}_m = 3.2$ m. Thus, the neglectable $0.1$ m improvement in $\bar{e}_m$ of \emph{$OUT^{w}$} over \emph{$OUT^{+}$} comes at 3the cost of a $20$\% higher energy consumption.

Now, instead of deriving a node-specific path loss exponent, $\eta_n$ in (\ref{eq:Pd}), to estimate the fade level of the link-channel pairs, we derive a global path loss exponent for all the nodes, and we still apply the link-channel pairs selection procedure in Section \ref{links_sel_weighting}. This approach was applied in \cite{MultiScale_2013}. We name this modified version of the outdoor RTI method \emph{$COM^{+}$}. With \emph{$COM^{+}$}, $11.7$\% of the link-channel pairs are selected, \emph{i.e.}, $\vartheta_e = 88.3$\%, and the false alarm rate is $0.06$\% (two triggered false alarms, person detected in the deployment area for $16$ s). The RMSEs are $\bar{e}_m = 5.0$ m and $\bar{e}_m = 4.2$ m. Thus, deriving a global path loss exponent decreases the localization accuracy of the system by approximately $30$\% compared to deriving one for each node.

Finally, in a version we name \emph{$COM^{w}$}, we still derive a global path loss exponent (as in \emph{$COM^{+}$}), but this time we apply the weighted average approach in (\ref{eq:weighted_RSSchange}) for estimating the change in RSS of the links. With \emph{$COM^{w}$}, the false alarm rate is $0.07$\% (two triggered false alarms, person detected in the deployment area for $17$ s), the percentage of selected link-channel pairs is $31.3$\%, \emph{i.e.}, $\vartheta_e = 68.7$\%, and the RMSEs are $\bar{e}_m = 4.9$ m and $\bar{e}_m = 4.1$ m. Results and characteristics of the outdoor RTI methods discussed above are summarized in Table \ref{t:RTI_methods_table}.

\subsection{Comparison with Previous RTI Methods} \label{comparison_results}

We now evaluate the performance of the RTI methods in \cite{MTT_Bocca_2013} and \cite{MultiScale_2013} on the same RSS measurements used to derive the results in Section \ref{new_methods_results}. Both methods were developed for time invariant indoor environments. Thus, in order to fairly evaluate their performance, we adapt both methods to nonstationary outdoor environments by applying the reference RSS estimation method in Section \ref{online_RSS_change} and the people detection and tracking method in Section \ref{MTT_subsection}.

Consider the method in \cite{MultiScale_2013}. The width $\lambda$ of the elliptical sensitivity area of the link-channel pairs of the network depends on both the fade level and the sign of the measured change in RSS. The fade level is estimated by deriving a global path loss exponent for all the nodes. The RSS variance of the link-channel pairs estimated in stationary conditions is not taken into consideration. We refer to this method as \emph{fade level-based}, or \emph{$FLB$}, and consider three different versions:

\begin{itemize}
    \item[\emph{$FLB^U$}:]all the link-channel pairs are selected.
    
    \item[\emph{$FLB^w$}:] the link-channel pairs with an average RSS, estimated in stationary conditions, lower than $\Upsilon_r$ are discarded. Among the remaining ones, only the link-channel pairs having positive fade level are selected. The weighted average approach in (\ref{eq:weighted_RSSchange}) is applied.
    
    \item[\emph{$FLB^+$}:] the link-channel pairs with an average RSS, estimated in stationary conditions, lower than $\Upsilon_r$ are discarded. For each link, only the frequency channel with the maximum fade level is selected.
\end{itemize}

The results are listed in Table \ref{t:indoor_RTI_methods}. Differently than with the outdoor RTI method, the localization accuracy of the \emph{$FLB$} method in \cite{MultiScale_2013} decreases when fewer link-channel pairs are selected. This makes the \emph{$FLB$} method less energy-efficient than the proposed outdoor RTI method. Moreover, even when all the link-channel pairs are considered, as with \emph{$FLB^U$}, the localization accuracy is approximately $20$\% lower than with \emph{$OUT^+$}, which uses only $12.8$\% of the link-channel pairs. In comparison, with \emph{$FLB^+$}, \emph{i.e.}, the most energy efficient version using only $16.9$\% of the link-channel pairs, the localization accuracy is approximately $125$\% worse than with the corresponding outdoor method \emph{$OUT^+$}. As far as the false alarm rate is concerned, the three considered versions of the method in \cite{MultiScale_2013} have a considerably worse performance than the proposed outdoor RTI method: the false alarm rate is $0.74$\% with \emph{$FLB^U$} ($12$ triggered false alarms, person detected in the monitored area for more than three minutes), and $1.41$\% with \emph{$FLB^+$} ($27$ triggered false alarms, person detected in the monitored area for approximately six minutes). These results demonstrate that the method in \cite{MultiScale_2013} is prone to trigger several false alarms in a time-varying and multipath-rich outdoor environment due to the significant noise in the RSS measurements introduced by environmental factors.

\begin{center}
\begin{table}[t!]
    \caption{Summary of the performance of previous RTI methods} 
        \centering
        \footnotesize
        \begin{tabular}{c | c c c c}
        \hline\hline 
        METHOD & $\vartheta_e$ & False alarm rate [\%] & $\bar{e}_m$ [m] & $\bar{e}_s$ [m]\\
        \hline
        \emph{$FLB^U$} & 0.0  & 0.74 & 4.3 & 4.1 \\
        \hline
        \emph{$FLB^w$} & 55.0 & 0.89 & 6.1 & 7.2 \\
        \hline 
        \emph{$FLB^+$} & 83.1 & 1.41 & 7.2 & 8.6 \\
        \hline 
        \emph{$RFL^p$} & 25.0 & 0.92 & 5.6 & 6.9 \\
        \hline 
        \emph{$RFL^f$} & 50.0 & 0.95 & 5.6 & 6.5 \\
        \hline 
        \emph{$RFL^+$} & 78.8 & 1.19 & 6.1 & 7.3 \\
        \hline 
        \end{tabular}
        \label{t:indoor_RTI_methods}
\end{table}
\end{center}

\begin{figure}[t!]
    \begin{center}
        \epsfig{figure=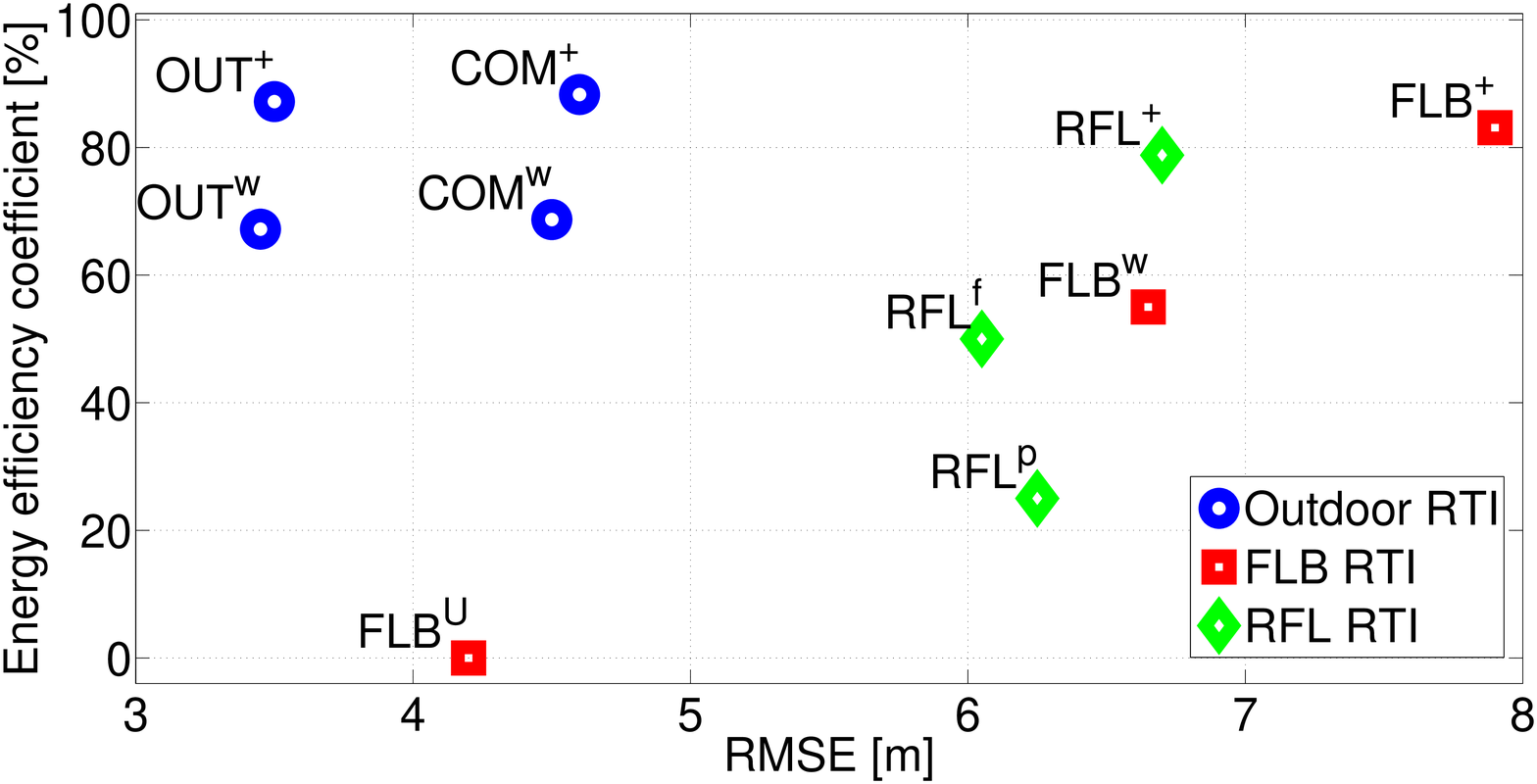,width=\columnwidth}
        \caption{Comparison of the performance of the considered RTI methods, in terms of localization accuracy and energy efficiency. The RMSE reported in the plot is computed as the average of $\bar{e}_s$ and $\bar{e}_m$.}
        \label{fig:results_comparison}
    \end{center}
\end{figure}

We now consider the RTI method in \cite{MTT_Bocca_2013}. There, the \emph{relative} fade level of a link-channel pair $\{l,c\}$ is estimated as the difference between its average RSS estimated in stationary conditions, $\bar{r}_{l,c}$ and the lowest average RSS measured on the frequency channels in $\mathcal{C}$:
\begin{equation} \label{eq:fade_level_MTT}
    F_{l,c} = \bar{r}_{l,c} - \min_{c \in \mathcal{C}} \bar{r}_{l,c}.
\end{equation}
Thus, $F_{l,c} \ge 0$, and $F_{l,c} = 0$ for one channel on each link. The RSS variance of the link-channel pairs estimated in stationary conditions is not taken into consideration. We refer to the RTI method in \cite{MTT_Bocca_2013} as \emph{relative fade level}, or \emph{RFL}, and consider three different versions:

\begin{itemize}
    \item[\emph{$RFL^p$}:]the link-channel pairs with $F_{l,c} > 0$ are selected. The change in RSS of the links is estimated by applying the weighted average approach in (\ref{eq:weighted_RSSchange}).
    
    \item[\emph{$RFL^f$}:]the link-channel pairs with an average RSS, measured in stationary conditions, lower than $\Upsilon_r$ are discarded. Among the remaining ones, the link-channel pairs with $F_{l,c} > 0$ are selected. The weighted average approach in (\ref{eq:weighted_RSSchange}) is applied.
    
    \item[\emph{$RFL^+$}:] the link-channel pairs with an average RSS, measured in stationary conditions, lower than $\Upsilon_r$ are discarded. For each link, only the frequency channel with the maximum fade level is selected.
\end{itemize}

The results are summarized in Table \ref{t:indoor_RTI_methods}. Despite selecting more link-channel pairs, the three considered versions of the method in \cite{MTT_Bocca_2013} have a localization accuracy from $73$\% (with \emph{$RFL^f$}) to $91$\% (with \emph{$RFL^+$}) worse than that of the most energy efficient outdoor RTI method \emph{$OUT^{+}$}. In addition, the method in \cite{MTT_Bocca_2013} has a significantly higher false alarm rate ($0.92$\% with \emph{$RFL^p$}, $0.95$\% with \emph{$RFL^f$}, $1.19$\% with \emph{$RFL^+$}) than the proposed outdoor RTI method ($0.04$\% with \emph{$OUT^+$}).

\subsection{Background Subtraction Contribution} \label{sec:back_sub_effect}

\begin{figure}[t!]
    \begin{center}
        \epsfig{figure=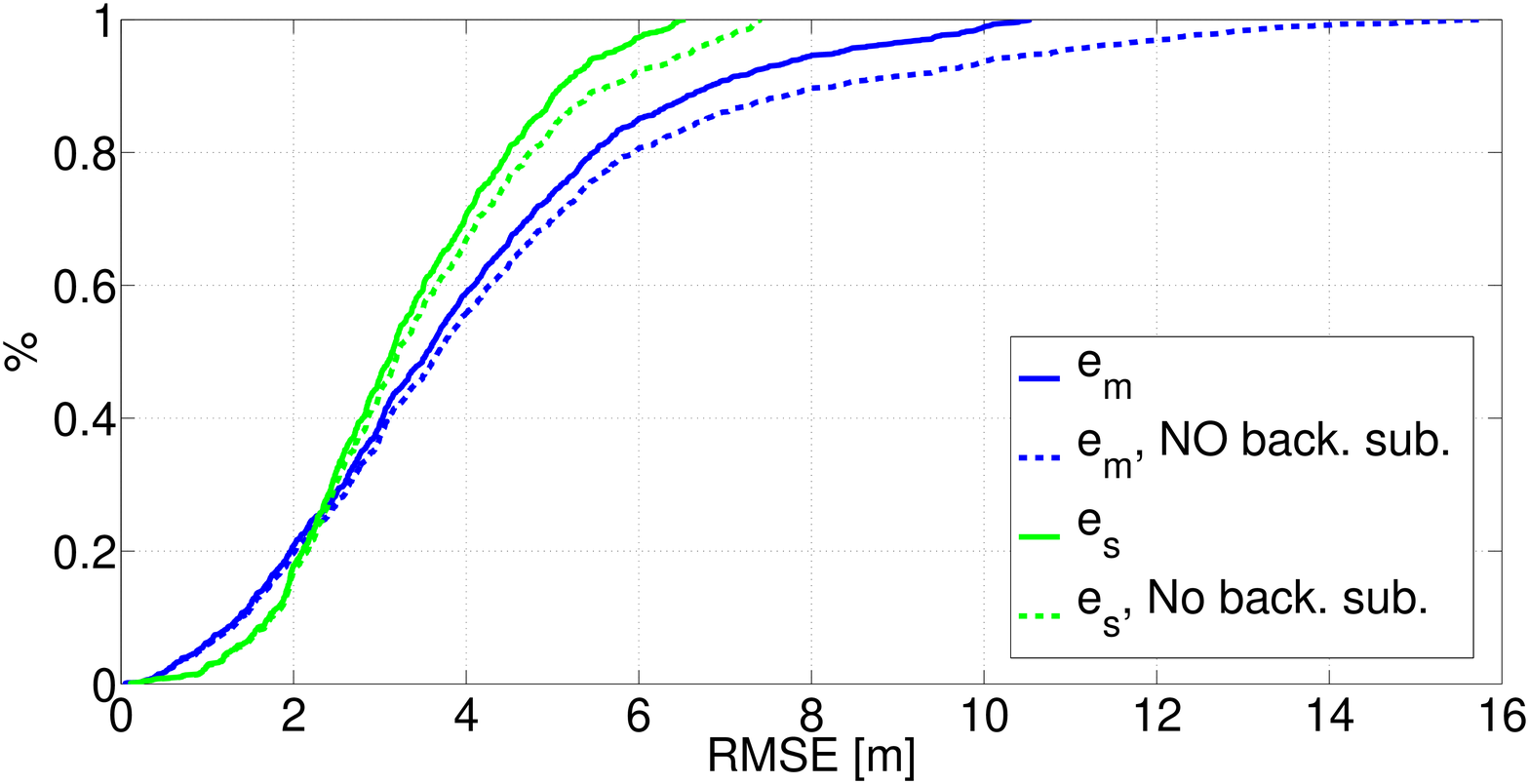,width=\columnwidth}
        \caption{Cumulative distribution functions of $\bar{e}_s$ and $\bar{e}_m$ of the novel outdoor RTI method ($OUT^{+}$), both when the background subtraction algorithm detailed in Section \ref{sec:background_sub} is applied (solid lines) or not applied (dashed lines).}
        \label{fig:back_sub}
    \end{center}
\end{figure}

We now analyze the effects of the background subtraction algorithm detailed in Section \ref{sec:background_sub} on the localization accuracy of the outdoor RTI method (\emph{$OUT^{+}$}). Figure \ref{fig:back_sub} shows the cumulative distribution functions of $\bar{e}_s$ and $\bar{e}_m$, both when background subtraction is applied (solid lines) or not applied (dashed lines). Background subtraction increases the robustness of the motion detection and localization process to the RSS variation due to environmental factors, which would introduce spurious blobs in the estimated radio tomographic images. Consequently, by applying the background subtraction algorithm, the RTI system is overall more accurate, and does not incur in very high (\emph{i.e.}, above $10$ m) localization errors when links located far away from the current position of the target are affected by a consistent environmental noise.

\begin{figure*}[t]
    \begin{center}
        \mbox{
            \subfigure[]{\epsfig{figure=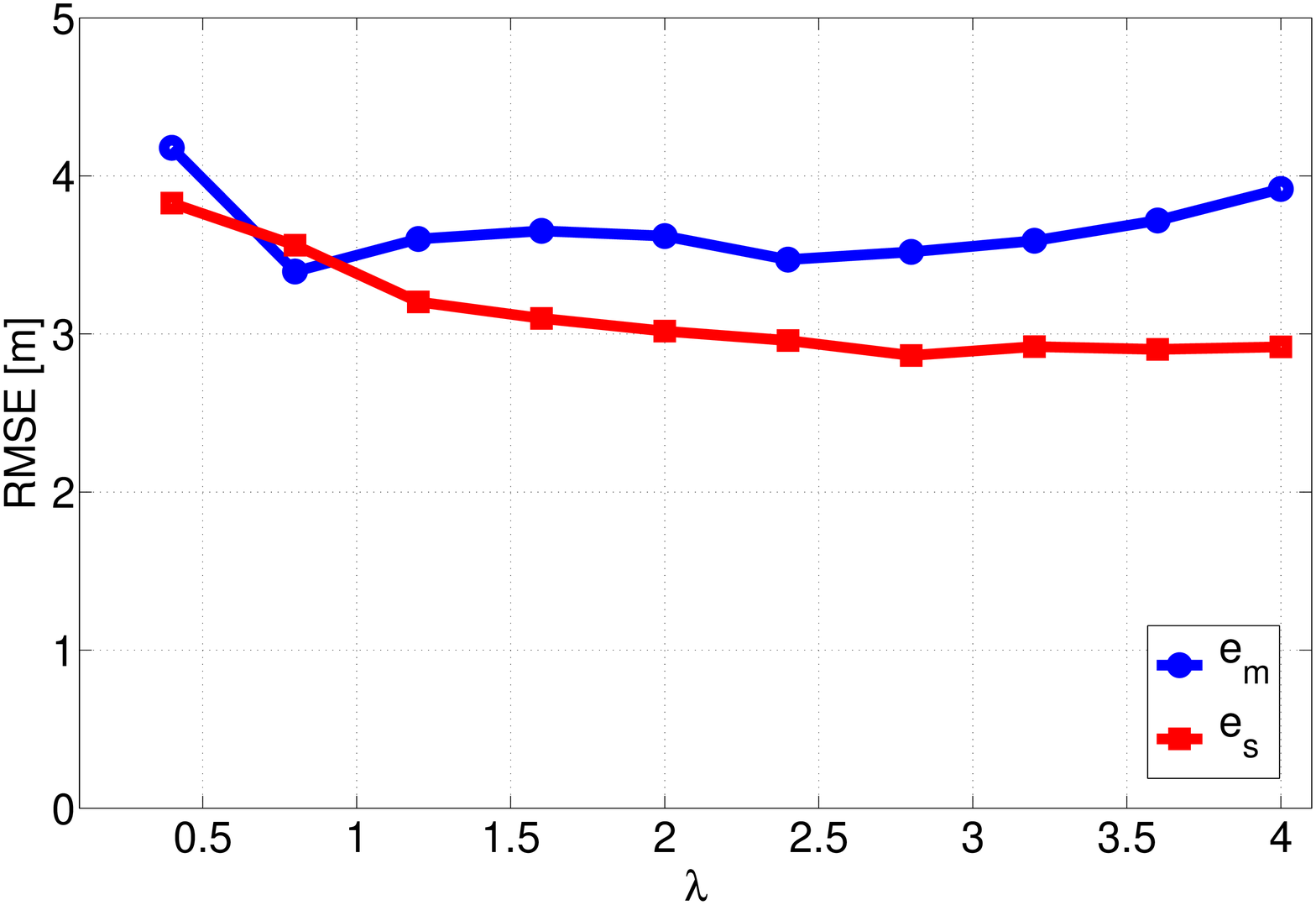,width=\columnwidth}} \quad
            \subfigure[]{\epsfig{figure=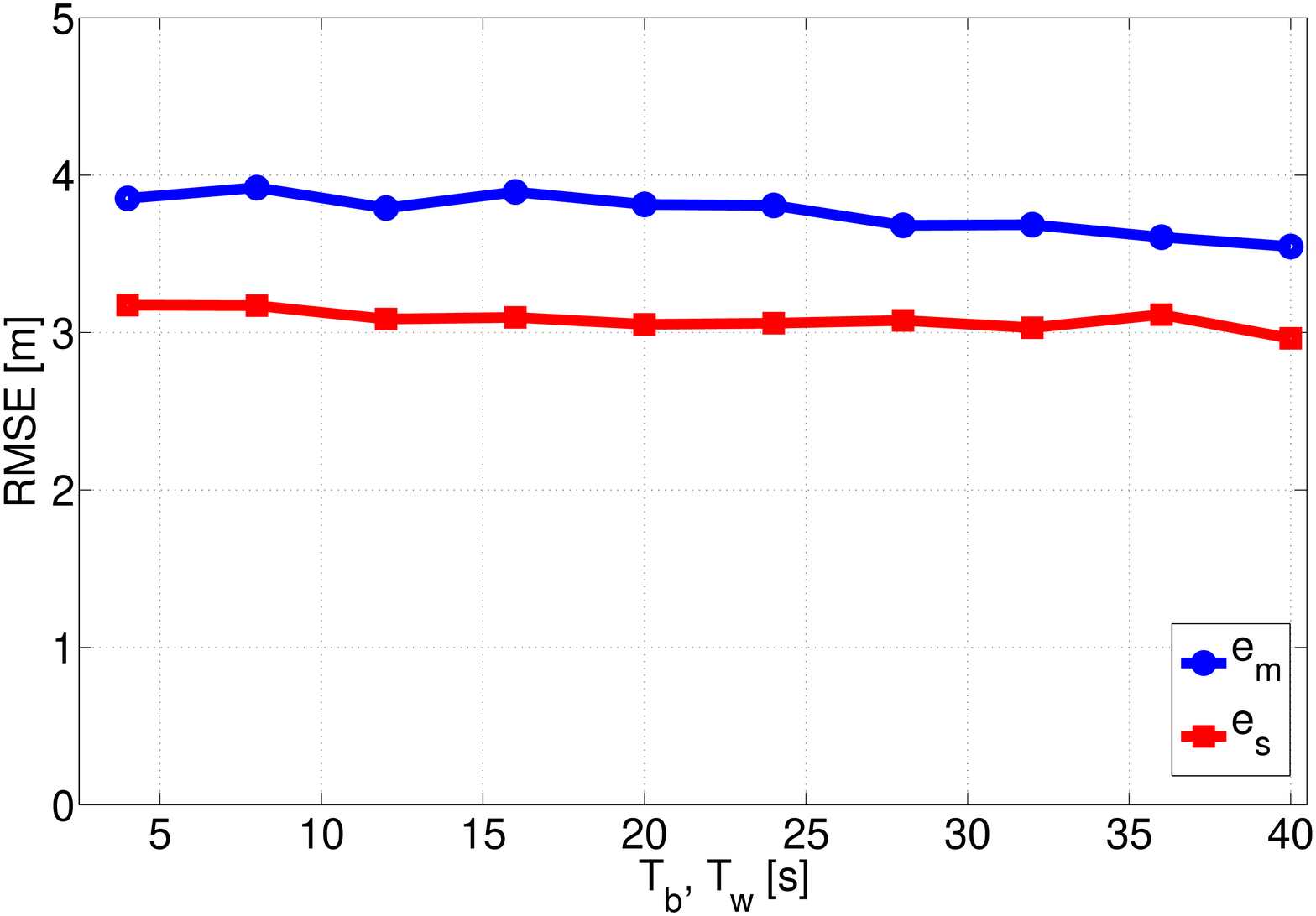,width=\columnwidth}}
        }
        \caption{Results of the sensitivity analysis of the energy-efficient outdoor RTI method (\emph{$OUT^{+}$}). In (a), the localization accuracy of the system with different values of $\lambda$, \emph{i.e.}, the width of the elliptical sensitivity area of the links. In (b), we modify the length of the time window used to determine the reference RSS of the links ($T_w$) and the length of the background image training period ($T_b$).}
        \label{fig:sens_analysis}
    \end{center}
\end{figure*}

\subsection{Sensitivity Analysis} \label{sens_analysis}

We analyze the effect of three parameters in Table \ref{t:table_parameters} on the localization accuracy of the system. Figure \ref{fig:sens_analysis} shows how $\bar{e}_m$ and $\bar{e}_s$ are affected by $\lambda$, \emph{i.e.}, the width of the elliptical sensitivity area of the links, $T_w$, \emph{i.e.}, the length of the time window used to determine the reference RSS of the links, and $T_b$, \emph{i.e.}, the length of the background image training period. Simulations are performed by post-processing the RSS measurements collected during the deployment of the system when the person was located inside the monitored area. In the simulations, we set $T_w = T_b$. The results in Figure \ref{fig:sens_analysis} indicate that the proposed outdoor RTI method is robust to variations of the three considered parameters, as both $\bar{e}_m$ and $\bar{e}_s$ have a $12$\% maximum variation from the values reported in table \ref{t:RTI_methods_table}.

\section{Conclusion} \label{sec:conclusion}

In this paper, we present and evaluate a novel RTI method for accurately detecting and tracking people in nonstationary outdoor environments. An RTI system is composed of static RF sensors, communicating on multiple frequency channels in the $2.4$ GHz ISM band, which continuously measure the RSS of the links of the mesh network to estimate the change in the electromagnetic field of the monitored area introduced by the presence and movements of people. The outdoor RTI method described in this work makes the system robust to the time-varying environmental noise typical of harsh outdoor environments, \emph{i.e.}, the variation in RSS introduced by environmental factors such as wind-driven foliage, rainfalls, or snow. In addition, the proposed method improves the energy efficiency of the system by selecting only link-channel pairs which are reliable indicators of the presence of a person on the link line. To experimentally verify the performance of the outdoor RTI method, we deploy a RTI system in a large forested area. The results demonstrate that the novel outdoor RTI method minimizes the false alarm rate and reduces significantly the localization error (from $20$\% to $46$\%) and energy consumption (from $62$\% to $87$\%) compared to already existing indoor RTI methods, which we suitably adapted to the considered outdoor scenario to grant a fair comparison.

In future works, we will focus on improving the energy efficiency of the RTI system by applying area coverage criteria to further reduce the set of selected link-channel pairs and by adaptively tuning the sampling frequency of the system based on the observed environmental conditions. In addition, we will investigate the robustness to environmental noise of other frequency bands (\emph{e.g.}, $433$ MHz and $900$ MHz).



\begin{thebibliography}{43}
\bibitem{RFSensNet_Proc_IEEE} N. Patwari and J. Wilson, "RF sensor networks for device-free localization and tracking: Measurements, models, and algorithms," Proceedings of the IEEE, vol. 98, no. 11, pp. 1961-1973, Nov 2010.

\bibitem{RTI_Wilson_TMC_2010} J. Wilson and N. Patwari, "Radio tomographic imaging with wireless networks," IEEE Transactions on Mobile Computing, vol. 9, no. 5, pp. 621-632, May 2010.

\bibitem{ChannelDiv_MASS_2012} O. Kaltiokallio, M. Bocca, and N. Patwari, "Enhancing the accuracy of radio tomographic imaging using channel diversity," 2012 IEEE Int'l Conference on Mobile Ad hoc and Sensor Systems (MASS 2012).

\bibitem{roadside_surv} C. Anderson, R. Martin, T. Walker, and R. Thomas, "Radio Tomography for Roadside Surveillance," IEEE Journal of Selected Topics in Signal Processing, vol. 8, no. 1, pp. 66-79, Feb 2014.
\bibitem{MTT_Bocca_2013} M. Bocca, O. Kaltiokallio, N. Patwari, and S. Venkatasubramanian, "Multiple Target Tracking with RF Sensor Networks", IEEE Transactions on Mobile Computing, vol 13, no. 8, pp. 1787-1800, Aug 2014.

\bibitem{Grandma_2012} O. Kaltiokallio, M. Bocca, and N. Patwari, "Follow @grandma: Long-term device-free localization for residential monitoring," 2012 IEEE Int'l Workshop on Practical Issues in Building Sensor Network Applications (SenseApp 2012).

\bibitem{srinivasan2006} K. Srinivasan, P. Dutta, A. Tavakoli, and P. Levis, "Understanding the causes of packet delivery success and failure in dense wireless sensor networks," 2006 Int'l Conference on Embedded Networked Sensor Systems (SenSys'06).

\bibitem{Basagni_NRGHarvesting} S. Basagni, M.Y. Naderi, C. Petrioli, and D. Spenza, "Wireless sensor networks with energy harvesting," Mobile Ad Hoc Networking: The Cutting Edge Directions, 2013, pp. 701-736.

\bibitem{drones_poachers} Reuters. "Kenya to use drones to fight elephant, rhino poachers," [Online]. Available: \url{http://www.reuters.com/article/2014/03/25/us-kenya-poaching-idUSBREA2O1D020140325}
\bibitem{Corke_VirtualFence} Z. Butler, P. Corke, R. Peterson, and D. Rus, "From robots to animals: Virtual fences for controlling cattle," Int'l Journal of Robotic Research, vol. 25, no. 5-6, pp. 485-508, May 2006.

\bibitem{wilson11fade} J. Wilson and N. Patwari, "A fade level skew-Laplace signal strength model for device-free localization with wireless networks," IEEE Transactions on Mobile Computing, vol. 11, no. 6, pp. 947-958, June 2012.

\bibitem{MultiScale_2013} O. Kaltiokallio, M. Bocca, and N. Patwari, "A fade level-based spatial model for radio tomographic imaging", IEEE Transactions on Mobile Computing, vol. 13, no. 6, pp. 1159-1172, June 2014.

\bibitem{RFIDs_system} X. Huang, R. Janaswamy, and A. Ganz, "Scout: Outdoor Localization Using Active RFID Technology," 2006 Int'l Conference on Broadband Communications, Networks and Systems (BROADNETS 2006).

\bibitem{CamerasRFID_system} R. Cucchiara, M. Fornaciari, A. Prati, and P. Santinelli, "Mutual calibration of camera motes and RFIDs for people localization and identification," 2010 ACM/IEEE Int'l Conference on Distributed Smart Cameras (ICDSC '10).
\bibitem{VisSensNet_2009} S. Soro and W. Heinzelman, "A survey of visual sensor networks," Advances in Multimedia, 2009.

\bibitem{Tahmoush_2009} D. Tahmoush and J. Silvious, "Remote detection of humans and animals," 2009 IEEE Applied Imagery Pattern Recognition Workshop (AIPRW'09).

\bibitem{Tahmoush_2013} D. Tahmoush and J. Silvious, "Radar measurement of human polarimetric {micro-Doppler}‚" Journal of Electrical and Computer Engineering, vol. 2013, Jan 2013.

\bibitem{MIMO_radar} R.O. Lane and S. Hayward, "Detecting personnel in wooded areas using MIMO radar," 2007 IET Int'l Conference on Radar Systems.

\bibitem{bannister2008wireless} K. Bannister, G. Giorgetti, and S. Gupta, "Wireless sensor networking for hot applications: Effects of temperature on signal strength, data collection and localization," 2008 Workshop on Embedded Networked Sensors (HotEmNets’ 08).
\bibitem{DISI_outdoor_2013} R. Marfievici, A. Murphy, G. Picco, F. Ossi, and F. Cagnacci, "How environmental factors impact outdoor wireless sensor networks: A case study," 2013 Int'l Conference on Mobile Ad-Hoc and Sensor Systems (MASS).

\bibitem{wind_radio_2006} M. Hashim and S. Stavrou, "Measurements and modelling of wind influence on radio wave propagation through vegetation," IEEE Transactions on Wireless Communications, vol. 5, no. 5, pp. 1055-1064, May 2006.

\bibitem{wilson10see} J. Wilson and N. Patwari, "See through walls: Motion tracking using variance-based radio tomography networks," IEEE Transactions on Mobile Computing, vol. 10, no. 5, pp. 612-621, May 2011.

\bibitem{Kaltiokallio_RTCSA_2011} O. Kaltiokallio and M. Bocca, "Real-time intrusion detection and tracking in indoor environment through distributed RSSI processing," 2011 IEEE Int'l Conference on Embedded and Real-Time Computing Systems and Applications (RTCSA 2011).

\bibitem{Zhao_kernel_2013} Y. Zhao, N. Patwari, J. Phillips, and S. Venkatasubramanian, "Radio tomographic imaging and tracking of stationary and moving people via kernel distance," 2013 IEEE/ACM Int'l Conference on Information Processing in Sensor Networks (IPSN '13).
\bibitem{Zhao_noise_2011} Y. Zhao and N. Patwari, "Noise reduction for variance-based device-free localization and tracking," 2011 IEEE Conference on Sensor, Mesh and Ad Hoc Communications and Networks (SECON 2011). 

\bibitem{kanso09b} M. A. Kanso and M. G. Rabbat, "Compressed RF tomography for wireless sensor networks: Centralized and decentralized approaches," 2009 IEEE Int'l Conference on Distributed Computing in Sensor Systems (DCOSS-09).

\bibitem{Khaledi_SECON_2014} M. Khaledi, S. Kasera, N. Patwari, and M. Bocca, "Energy efficient radio tomographic imaging," 2014 IEEE Int'l Conference on Sensing, Communication, and Networking (SECON 2014).

\bibitem{patwari08b} N. Patwari and P. Agrawal, "Effects of correlated shadowing: Connectivity, localization, and RF tomography," 2008 IEEE/ACM Int'l Conference on Information Processing in Sensor Networks (IPSN'08).

\bibitem{agrawal09} P. Agrawal and N. Patwari, "Correlated link shadow fading in multi-hop wireless networks," IEEE Transactions on Wireless Communications, vol. 8, no. 8, pp. 4024-4036, August 2009.
\bibitem{rappaport96} T. Rappaport, "Wireless Communications: Principles and Practice," New Jersey: Prentice-Hall Inc., 1996.

\bibitem{HW_RSS_var_2006} D. Lymberopoulos, Q. Lindsey, and A. Savvides, "An empirical characterization of radio signal strength variability in 3-d IEEE 802.15.4 networks using monopole antennas," 2006 European Conference on Wireless Sensor Networks (EWSN'06).

\bibitem{Vallet_model} J. Vallet, O. Kaltiokallio, M. Myrsky, J. Saarinen, and M. Bocca, "Simultaneous RSS-based localization and model calibration in wireless networks with a mobile robot," 2012 Int'l Conference on Ambient Systems, Networks and Technologies (ANT 2012).

\bibitem{RSSI_underappreciated} K. Srinivasan and P. Levis, "RSSI is under appreciated," 2006 Workshop on Embedded Networked Sensors (EmNets).

\bibitem{Ceriotti_jungle} M. Ceriotti, M. Chini, A.L. Murphy, G.P. Picco, F. Cagnacci and B. Tolhurst, "Motes in the jungle: Lessons learned from a short-term WSN deployment in the Ecuador cloud forest," 2010 Int'l Workshop on Real-World Wireless Sensor Networks (RealWSN'10).
\bibitem{Vogel_book} C.R. Vogel, "Computational methods for inverse problems," SIAM, 2002.

\bibitem{Piccardi_back_sub} M. Piccardi, "Background subtraction techniques: a review," 2004 Int'l Conference on Systems, Man and Cybernetics.

\bibitem{Benezeth_back_sub} Y. Benezeth, P.-M. Jodoin, B. Emile, H. Laurent, and C. Rosenberger, "Review and evaluation of commonly-implemented background subtraction algorithms," 2008 Int'l Conference on Pattern Recognition (ICPR 2008).

\bibitem{TI_CC2531} Texas Instruments. A USB-Enabled system-on-chip solution for 2.4 GHz IEEE 802.15.4 and ZigBee applications. [Online]. Available: \url{http://www.ti.com/lit/ds/symlink/cc2531.pdf}

\bibitem{antenna_nodes} Texas Instruments. 2.4 GHz Inverted F Antenna. [Online]. Available: \url{http://www.ti.com/lit/an/swru120b/swru120b.pdf}
\bibitem{RTI_book} M. Bocca, O. Kaltiokallio, and N. Patwari, "Radio tomographic imaging for ambient assisted living," Springer - Communications in Computer and Information Science, vol. 362, pp. 108-130, 2013.

\bibitem{GPS_forest} P. Sigrist, P. Coppin, and M. Hermy, "Impact of forest canopy on quality and accuracy of GPS measurements," Int'l Journal of Remote Sensing, vol. 20, no. 18, pp. 3595-3610, 1999.

\bibitem{GPS_forest2} A.S. Bastos and H. Hasegawa, "Behavior of GPS signal interruption probability under tree canopies in different forest conditions," European Journal of Remote Sensing, vol. 46, pp. 613-622, Oct. 2013.

\bibitem{Alippi_book} C. Alippi, "Intelligence for Embedded Systems: A Methodological Approach," 1st ed. Springer, 2014.


\end{thebibliography}
\end{document}